\documentclass[twocolumn,showpacs,preprintnumbers,amsmath,amssymb]{revtex4}

\usepackage[usenames,dvipsnames]{xcolor} 
\usepackage[latin1]{inputenc}
\usepackage[T1]{fontenc}
\usepackage{latexsym}
\usepackage{amsfonts}
\usepackage{amssymb}
\usepackage{amsmath}
\usepackage{psfrag}
\usepackage{rotating}
\usepackage{pgf,pgfsys,pgffor}
\usepackage{pgfplots}
\usepackage{pgfplotstable}
\usepackage{hyperref}

\def\diag{{\rm diag}}

\def\E[#1]{{\rm E}\left[#1\right]}
\def\Ec[#1|#2]{{\rm E}\left[#1|#2\right]}

\def\lb{\lambda}
\def\M(#1){\mathbb{#1}}
\def\mn{\medskip\noindent}

\def\P[#1]{{\rm P}\left[\,#1\,\right]}
\def\Pc[#1|#2]{{\rm P}\left[#1|#2\right]}

\usepackage{
amssymb
}
\usepackage{graphicx} 

\usepackage{color}

\newcommand{\ket}[1]{\vert {#1} \rangle}

\newcommand{\beq}{\begin{equation}}
\newcommand{\eeq}{\end{equation}}

\input alli

\def\b(#1){\langle#1|}
\def\B(#1){{\bf #1}}
\def\C(#1){{\cal #1}}

\def\diag{{\rm diag}}
\def\E[#1]{{\rm E}\left[#1\right]}
\def\Ec[#1|#2]{{\rm E}\left[#1|#2\right]}

\def\k(#1){|#1\rangle}
\def\bk(#1,#2){\langle#1|#2\rangle}
\def\bok(#1,#2,#3){\langle#1|#2|#3\rangle}
\def\kb(#1,#2){|#1\rangle\langle#2|}
\def\lb{\lambda}
\def\M(#1){\mathbb{#1}}
\def\mn{\medskip\noindent}

\def\P[#1]{{\rm P}\left[\,#1\,\right]}
\def\Pc[#1|#2]{{\rm P}\left[\,#1\,|\,#2\,\right]}
\def\tit(#1){``#1''}
\def\tgamma{\tilde\gamma}
\def\Tr{{\rm Tr}\,}
\input alli

\begin{document}

\title{On the Optimality of Square Root Measurements in Quantum State Discrimination}

%

\author{Nicola Dalla Pozza}
  \email{n.dallapozza@adfa.edu.au}
  \affiliation{School of Engineering and Information Technology, University of New South Wales Canberra, Australia}

\author{Gianfranco Pierobon}%
  \affiliation{University of Padova, Padova, Italy}%
  
\date{\today}

%

\begin{abstract}
Distinguishing assigned quantum states with assigned probabilities via
quantum measurements is a crucial problem for the transmission of
classical information through quantum channels. Measurement operators
maximizing the probability of correct discrimination have been
characterized by Helstrom, Holevo and Yuen since 1970's. On the other
hand, closed--form solutions are available only for particular
situations enjoying high degrees of symmetry. As a suboptimal
solution to the problem, measurement operators, directly determined
from states and probabilities and known as square root measurements
(SRM), were introduced by Hausladen and Wootters. These operators
were also recognized to be optimal for pure states equipped with
geometrical uniform symmetry (GUS). In this paper we discuss the
optimality of the SRM and find necessary and sufficient conditions
in order that SRM maximize the correct decision probabilities for
set of states formed by several constellations of GUS states. The
results are applied to some specific examples concerning double
constellations of quantum phase shift keying (PSK) and pulse
position modulation (PPM) states, with possible applications to
practical systems of quantum communications.
\end{abstract}


\maketitle

\section{\label{introduction}Introduction}

The long standing problem of transmitting classical information
through a noiseless quantum channel refers to the following
scenario. A classical information source emits a symbol $S$ in a
finite alphabet, say $\{s_0,\ldots,s_{m-1}\}$, whose nature is
irrelevant to the problem.  The symbols $s_i$ are emitted with
probabilities $q_i$ satisfying the properties $q_i>0$ and $\sum_{
i=0}^{m-1}q_i=1$. On the basis of the symbol $s_i$ emitted by the
classical source, the transmitter (conventionally known as Alice)
prepares the quantum channel in a state belonging to a set of agreed
quantum states, generally mixed states with density operators
$\rho_i$ which are semidefinite positive ($\rho_i\ge0$) and have
unitary trace ($\Tr\rho_i=1$). The quantum ensemble $\{q_i,\rho_i\}$
can be conveniently represented by the weighted density  operators
$W_i=q_i\rho_i$.

The problem of the receiver (conventionally known as Bob) is to
extract as well as he can classical information from the quantum
channel he and Alice share. In this task Bob is assumed to know both
the quantum states $\rho_i$ and their probabilities $q_i$. We assume
that the measurement and decision strategy of Bob is the {\it
quantum hypothesis testing} introduced by Helstrom \cite{Helstrom1976}. The
strategy consists in performing on the quantum channel a  positive
operator--valued measurement (POVM) with measurement operators $P_i$
semidefinite positive ($P_i\ge0$) and resolving the identity
operator, namely, $\sum_{i=0}^{m-1}P_i=I$. The conditional
probability $p\,(j|i)$ that the measurement gives the result $R=j$
provided that the state is $\rho_i$ turns out to be $\Tr(\rho_i
P_j)$. The usual task of Bob is the  choice of the POVM  $P_i$ {\it
maximizing the probability of correct decision}
\beq
P_c=\sum_{i=0}^{m-1}q_ip\,(i|i)=\sum_{i=0}^{m-1}q_i\Tr(\rho_iP_i)
	=\sum_{i=0}^{m-1}\Tr(W_iP_i)\;.
\label{defPc}
\eeq
The problem solution is given by well known general results due to
Holevo \cite{Holevo1973b}, Yuen {\it et al.} \cite{Yuen1975}, and Helstrom
\cite{Helstrom1976} (see also \cite{Barnett2009}, and \cite{Konig2009} for a connection with min--entropy of a classical source) and is summarized by the
following theorem.

\mn{\sc Theorem 1.} The measurement operators $P_i$ give the
maximum correct detection probability if and only if the operator
\beq
Y=\sum_{i=0}^{m-1}W_iP_i
\label{lagrangianoY}
\eeq
satisfies the conditions
\beq
Y-W_i\ge0
\label{condizioniOttimeSdp}
\eeq
for each $i=0,\ldots,m-1$. The above condition implies also that $Y$
is semidefinite positive and that the equalities
\beq
P_i(W_i-W_j)P_j=0
\label{condizioniOttimeUguaglianzaZero}
\eeq
hold true for all $i,j=0,\ldots,m-1$.$\hfill\square$

\medskip

Unfortunately, closed form solutions are available only in few very
particular cases, generally implying pure states enjoying high
degrees of symmetry. Indeed, as pointed out in \cite{Eldar2003}, serious
mathematical difficulties arise because of the non linearity of the
problem, which turns out to be a convex semidefinite programming
problem.

If the states prepared by Alice are pure, namely, $\rho_i=\kb(
\gamma_i,\gamma_i)$, it can be shown \cite{Belavkin1975} that also the
measurement operators have rank 1, namely, $P_i=\kb(\mu_i,\mu_i)$.
Then the correct detection probability \eqref{defPc} depends on the inner
products of the states $\k(\gamma_i)$ and of the {\it measurement
vectors} $\k(\mu_i)$, namely,
\beq
P_c=\sum_{i=0}^{m-1}|\bk(\mu_i,\tgamma_i)|^2\;,
\label{defPc2}
\eeq
where $\k(\tgamma_i)=\sqrt{q_i}\k(\gamma_i)$ are the weighted states.
The optimization problem reduces to find the measurement vectors
$\k(\mu_i)$ maximizing the correct detection probability \eqref{defPc2}. In
accordance with the general case, the vectors $\k(\mu_i)$ must
resolve the identity operator $I$ of the Hilbert space spanned by
the states
\beq
\sum_{i=0}^{m-1}\kb(\mu_i,\mu_i)=I\;.
\label{resolutionIdentity}
\eeq

In the following we assume that the weighted states $\k(\tgamma_i)$
are linearly independent. This is not at all too restrictive for all
situations of practical interest. In this case it can be shown
\cite{Kennedy1973b} that the optimal measurement vectors $\k(\mu_i)$ (to be
determined) form an orthonormal basis.  Since the weighted states
are linear combinations of the measurement vectors, provided that
the states $\k(\tgamma_i)$ are collected  into a matrix $\Gamma$ and
the measurement vectors $\k(\mu_i)$ into  a matrix $M$, i.e.,
\beq
\Gamma=[\,\k(\tgamma_0),\ldots,\k(\tgamma_{m-1})\,], \quad
M=[\,\k(\mu_0),\ldots,\k(\mu_{m-1})\,]
\eeq
a matrix relation $\Gamma=MX$ holds. The entries of the $m\times m$
matrix $X$ are given by $X_{ki}=\bk(\mu_k,\tilde\gamma_i)$. The joint
input--output probabilities become $p\,(i,j)=\P[S=i,R=j]=|X_{ij}|^2$
and the corresponding correct decision probability turns out to be
\beq
P_c=\sum_{i=0}^{m-1}|X_{ii}|^2\;.
\label{defPcWithX}
\eeq

The optimization problem can be conveniently formulated introducing
the {\it Gram matrix} $G$ of the weighted states $\k(\tgamma_i)$,
namely $G_{ij}=\bk(\tgamma_i,\tgamma_j)$. Then, using \eqref{resolutionIdentity} gives
\beq
G_{ij}=\sum_{k=0}^{m-1}\bk(\tilde\gamma_i,\mu_k)\bk(\mu_k,\tilde
	\gamma_j)	=\sum_{k=0}^{m-1}X_{ki}^*X_{kj}
	\label{gramMatrixWithX}
\eeq
and $G=X^\dagger X$ in matrix form. In conclusion, the optimal
detection problem reduces to find the factorization $G=X^\dagger X$
of the Gram matrix that maximizes probability \eqref{defPcWithX}. Provided that
$X_0$ is the result of the optimization, the measurement vectors are
obtained by $M=\Gamma X_0^{-1}$. Note that the independence of the
states implies that the Gram matrix is positive definite, so that $X$
is invertible. Unfortunately the factorization \eqref{gramMatrixWithX} of the Gram
matrix leads to a set of quadratic equations, which can be solved
only by numerical programs, at least in general.

A different approach is based on a very popular suboptimal
measurement, known as {\it square root measurement} (SRM), which was
introduced by Hausladen and Wootters \cite{Hausladen1994} and thoroughly
discussed by Eldar and Forney \cite{Eldar2001}. In the framework
discussed above, the square root measurement corresponds to the
factorization $G=X^\dagger X$ of the Gram matrix with $X=G^{1/2}$
square root of the Gram matrix itself. Under our assumptions both
the matrix $G^{1/2}$ and its inverse $G^{-1/2}$ are definite
positive. The matrix $M$ of the measurement vectors is given by $M=
\Gamma G^{-1/2}$ and the correct detection probability is the sum of
the squares of the diagonal entries of $G^{1/2}$, i.e.,
\beq
P_c=\sum_{i=0}^{m-1}[\,(G^{1/2})_{ii}\,]^2\;.
\label{defPcWithG}
\eeq
%

Even though this measurement is not optimal in general, it exhibits
interesting properties. It can be straightforwardly obtained by the
Gram matrix and gives performances near to the optimum provided that
the weighted states $\k(\tgamma_i)$ are almost orthogonal. 
Indeed Hausladen {\it et al.} \cite{Hausladen1996} have shown that SRM (also known as ''pretty good''
measurement) is asymptotically optimal, in the sense that it is good enough to be 
used as non local measurement in the proof of their fundamental theorem on the
classical capacity of a quantum channel. More recently SRM  has found application
\cite{Bacon2005, Moore2007, Hayashi2008} in the search of optimal measurement for distinguishing hidden subgroup 
states, providing an interesting approach to the large class of quantum 
computational problems linked to the {\it hidden subgroup problem}. In any case
SRM furnishes a lower bound to the
optimal performance and it may be assumed as a starting point to
find the optimal measurement $X_0$ through unitary transformations.

The SRM turns out to be the optimal measurement when the states
exhibit a high degree of symmetry. Then, particular attention has
been given to applications of the SRM to states
enjoying {\it geometrically uniform simmetry} (GUS), i.e., sets of
states that are invariant with respect to a unitary transformation
\cite{Ban1997,Eldar2001}. This is particularly interesting for
practical quantum communication systems using quantum phase shift--
keying (QPSK) \cite{Kato1999} and pulse position modulation (PPM) \cite{Cariolaro2010}. Also extensions to mixed state have been considered \cite{Eldar2004,Cariolaro2010b}.

The aim of the present paper is to show that the optimal feature of
SRM is by no means confined to states enjoying GUS and having equal
probabilities. In particular we consider a set of states composed by
distinct {\it constellations} of states enjoying the same GUS (we use the term ``constellation'' borrowed from the telecommunications 
jargon  for the more general ``subset''). This is a
particular case of the {\it compound geometrical uniform symmetry}
discussed in \cite{Eldar2004}. The novelty of our approach stays in the
fact that we find necessary and sufficient conditions in order that
SRM is the optimal measurement for this case. Moreover we present
possible applications to practical quantum communication systems as
PSK and suggest a version of PPM improving its efficiency.

The paper is organized as follows. In Section II we revisit two
important results concerning our topic. The first, due to Helstrom \cite{Helstrom1982}, characterizes the factorization $G=X^\dagger X$ that
maximixes the correct decision probability. The second, due to
Sasaki {\it et al.} \cite{Sasaki1998}, gives simple sufficient conditions
guaranteeing that the SRM is the optimal measurement. In particular
the last result enables one to show in a very simple way that SRM is
optimal for states $\k(\gamma_i)$ enjoying {\it geometrical uniform
symmetry}. The main result of the paper is presented in Section
III, where a necessary and sufficient condition is furnished in
order that the SRM is optimal for a set formed by constellations of
GUS states generated by the same unitary transformation $S$ but
applied to different states. In Section IV the result is applied to
a double constellation of quantum PSK states, showing that SRM may
be optimal also for states with non uniform probabilities. In Section
V we consider a double constellation of PPM states and show that the
SRM is optimal for this case and leads to a communications scheme
more efficient than the original PPM. Some conclusions close the
paper.

\section{The SRM as optimal measurement}

While Theorem 1 refers to the general case of mixed states, a simple 
characterization of the optimal measurement for independent pure states is
given by the following theorem.

\mn{\sc Theorem 2.} The factorization $G=X^\dagger X$ maximizes the
correct decision probability if and only if the following conditions
hold:

i)  for each $i$ and $j$
\beq
X_{ii}X_{ji}^*=X_{ij}X_{jj}^*\;;
\label{conditionHermitianity}
\eeq

ii) the matrix
\beq
Y=XX_d^\dagger
\label{conditionHelstrom}
\eeq
with
\beq
X_d=\diag\{X_{00},\ldots,X_{m-1,m-1}\}
\eeq
is positive definite. (Note that \eqref{conditionHermitianity} implies that $Y$ is Hermitian).
$\hfill\square$

\medskip

The theorem, not frequently cited in the literature, has been proved
by Helstrom in a paper \cite{Helstrom1982} concerning an iterative search
of the optimal measurement. The proof of the theorem, given in
appendix of the paper, is somewhat intricate. A simplified version
of the proof is given here in Appendix A.

As we pointed out, the SRM corresponding to the factorization $G=
X^\dagger X$ with $X=G^{1/2}$ is not optimal in general. 
On the basis of the Theorem 2, Sasaki {\it et al.} \cite{Sasaki1998} in a paper about the 
superadditivity of the capacity of a quantum channel have found a nice sufficient
condition for the optimality of SRM.
The following
theorem gives a generalization of the result.

\mn{\sc Theorem 3}. Assume that the Gram matrix $G$ (and its square
root $G^{1/2}$) is block diagonal, namely, $G=G_1\oplus\ldots\oplus
G_r$. Then, the square root measurement is optimal if and only if
the square root $G_i^{1/2}$ of each block has equal diagonal entries.

\mn{\sc Proof.} If the matrix $G$ is formed by a single block, it is
irreducible and its associated directed graph is strongly connected \cite{HornJohnson}.
Then, the indexes $0,\ldots,m-1$ may be ordered into a cycle $i_0,i_1
,\ldots,i_{m-1},i_0$ such that $(G^{1/2})_{i_ki_{k+1}}\neq0$ for
each $k=0,\ldots,m-1$. Since $G^{1/2}$ is Hermitian, $(G^{1/2})_{
i_{k+1}i_k}^*=(G^{1/2})_{i_ki_{k+1}}$ and condition \eqref{conditionHermitianity} is satisfied
if and only if $(G^{1/2})_{i_ki_k}=(G^{1/2})_{i_{k+1}i_{k+1}}$ for
each $k$ or, equivalently, if the diagonal entries of $G^{1/2}$ have
a common value $g$. Therefore, condition \eqref{conditionHelstrom} of Theorem 2 holds in
that $XX_d^*=gG^{1/2}$ is positive definite. If $G$ has many diagonal
block, the proof can be applied to each block.$\hfill\square$

\medskip

Confining our attention to the case of a single block, since $Y_{ii}=
|X_{ii}|^2=p\,(i,i)$, the theorem is equivalent to say that the SRM
is optimal if and only if the probabilities of correct decision is
$g^2$, independent of the state transmitted. In particular, the
optimal correct decision probability turns out to be
\beq
P_c=mg^2\;.
\eeq

\medskip

It is worthwhile to note that the optimality of the SRM depends only
on the Gram matrix. Since this is invariant with respect to unitary
transformations, if a constellation of weighted states satisfy the
conditions of  Theorem 3, any other constellation obtained via a
unitary transformation admits optimal SRM with the same error
probability. Of course the optimal SRM operators vary according to
the unitary transformation.

The most known case of optimality of the SRM concerns states equipped
with {\it geometrical uniform symmetry} (GUS) \cite{Ban1997,Eldar2001}. 
The weighted states $\k(\tgamma_i)$ enjoy GUS if $\k(\tgamma_i)=S^i\k(\tgamma_0)$,
$i=0,\ldots,m-1$, where $S$ is a unitary operator satisfying the condition $S^m=I$.
This generalizes the usual definition of GUS states $\k(\gamma_i)$ as  satisfying
the condition $\k(\gamma_i)=S^i\k(\gamma_0)$, and having equal probabilities 
$q_i=1/m$.
The proof of the optimality of the SRM is an
immediate consequence of Theorem 3. Indeed, 
the entries of the weighted Gram matrix become
\beq
G_{ij}=\bk(\tgamma_i,\tgamma_j)=\bok(\tgamma_0,S^{j-i},\tgamma_0)
\eeq
and depend only by $(j-i)_{{\rm mod}\ m}$. One concludes that the
matrix $G$ is circulant \cite{Davis1979}. (Details on the basic
properties of circulant matrices are collected for convenience in
Appendix B). Then $G$ has spectral decomposition
\beq
G=F\Lambda F^\dagger
\eeq
(see \eqref{B3}) where $F$ is the unitary Fourier matrix \eqref{B4} and the
diagonal matrix $\Lambda=\diag\{\lb_1,\ldots,\lb_m\}$ collects the
(positive) eigenvalues of $G$. Then
\beq
G^{1/2}=F\Lambda^{1/2}F^\dagger
\eeq
is circulant, its diagonal entries are equal, and the SRM is
optimal. In particular
\beq
g=\frac{1}{m}\Tr(G^{1/2})=\frac{1}{m}\Tr(\Lambda^{1/2})=\frac{1}{m}
	\sum_{i=1}^m\lb_i^{1/2}
\eeq
and
\beq
P_c=mg^2=\frac{1}{m}
	\left(\sum_{i=1}^m\lb_i^{1/2}\right)^2\;.
\label{defPcWithLambda}
\eeq

The GUS is by no means the only case where Theorem 3 finds
application. As an elementary example consider the Gram matrix
\beq
G=\frac{1}{2}\left[\begin{matrix}1&\chi\cr \chi^*&1\cr \end{matrix}\right]
\eeq
of two states $\k(\gamma_0)$ and $\k(\gamma_1)$ with equal
probability and $\chi=\bk(\gamma_0,\gamma_1)$. Note that this matrix
is not circulant in general. Its square root matrix is
\beq
G^{1/2}=\frac{1}{\sqrt{2}}\left[\begin{matrix}a&b\cr b^*&a\cr\end{matrix}\right]
\eeq
with $a^2+|b|^2=1$ and $2ab=\chi$. By virtue of Theorem 3 the SRM is
optimal. A simple algebra shows that
\beq
a=\frac{1}{\sqrt{2}}\sqrt{1+\sqrt{1-|\chi|^2}}\;,
\eeq
so that the {\it Helstrom bound}
\beq
P_c=\frac{1}{2}\left(1+\sqrt{1-|\chi|^2}\right)
\eeq
holds. This is perhaps the simplest proof of the Helstrom bound, at
least for equilikely states.
In the literature the SRM is usually presented in connection with GUS states having 
equal probabilities. However, the extension to other situations is possible (see \cite{Mochon2006}
for a detailed, but scarcely constructive, theoretical analysis). Recently Kato \cite{Kato2013}
has found a closed form expression of the input probabilities for which the measurement operators 
obtained from the not weighted Gram matrix are optimal
in the case of three coherent states $\k(-\alpha)$, $\k(0)$ and $\k(\alpha)$.

\section{Multiple constellations of GUS states}

Here we consider a collection of $sm$ states $\{\k(\gamma_{
ki})\}$, $k~=~1,\ldots,s$, $i~=~0,\ldots,m-1$, defined as
\beq
\k(\gamma_{ki})=S^i\k(\gamma_{k0})\;,
\eeq
where $S$ is a unitary operator such that $S^m=I$. In other words
the set of states is formed by $s$ constellations of GUS states
obtained by the same unitary operator $S$, starting from
$s$ different states $\k(\gamma_{k0})$. Moreover we relax the
hypothesis of equal probabilities, only assuming that the states of
each constellation  have the same probability, namely, $q_{ki}=q_k$
with $\sum_kq_k=1/m$. Under these assumptions we find a sufficient
and necessary condition for the optimality of the SRM. This is a
particularization of the concept of {\it compound geometrical
uniform} (CGU) states, introduced by Eldar {\it et al.}
\cite{Eldar2004}, where $\k(\gamma_{ki})=S_i\k(\gamma_{k0})$ and the $m$
unitary operators $S_i$ form a (not necessarily Abelian)
multiplicative group.

We begin by ordering the states as
\beq
\k(\gamma_{10}),\ldots,\k(\gamma_{1,m-1}),\ldots,\k(\gamma_{s0}),
\ldots,\k(\gamma_{s,m-1})\;.
\label{orderingStates}
\eeq
The weighted Gram matrix of the states may be partitioned in the form
\beq
G=\left[\begin{matrix}G_{11}&\ldots&G_{1s}\cr
							\vdots&\ddots&\vdots\cr
							G_{s1}&\ldots&G_{ss}\cr\end{matrix}\right]\;,
\eeq
where the $s^2$ submatrices $G_{hk}$ have order $m$ and entries
\begin{align*}
G_{hk}(i,j)& =\bk(\tgamma_{hi},\tgamma_{kj})=\sqrt{q_hq_k}
	\bk(\gamma_{hi},\gamma_{kj}) \nonumber \\
	& =\sqrt{q_hq_k}\bok(\gamma_{h0},S^{j-i},\gamma_{k0})\;.
\end{align*}
It follows that the matrices $G_{hk}$ are circulant and are
simultaneously diagonalizable as $G_{hk}=F\Lambda_{hk}F^\dagger$,
where $F$ is the Fourier matrix defined in Appendix B. Then $G=F_s
\Lambda F_s^\dagger$, with $F_s=F\oplus\ldots\oplus F$ and the
blocks of
\beq
\Lambda=\left[\begin{matrix}\Lambda_{11}&\ldots&\Lambda_{1s}\cr
							\vdots&\ddots&\vdots\cr
							\Lambda_{s1}&\ldots&\Lambda_{ss}\cr\end{matrix}\right]
\label{defLambda}
\eeq
are diagonal matrices of order  $m$.

The square root of $\Lambda$, say  $\Sigma=\Lambda^{1/2}$, has the
same structure as $\Lambda$. Indeed, by a joint permutation of rows
and columns generated by a permutation matrix $\Pi$ the matrix $
\Lambda$ can be transformed into a block diagonal matrix
\beq
D=\Pi\Lambda\Pi^T=\left[\begin{matrix}D_1&\ldots&0\cr
					\vdots&\ddots&\vdots\cr0&\ldots&D_s\cr\end{matrix}\right]\;.
\label{defD}
\eeq
Incidentally, the permutation converting the ordering \eqref{orderingStates} of the
states into the ordering
\beq
\k(\gamma_{10}),\ldots,\k(\gamma_{s0}),\ldots,\k(\gamma_{1,m-1}),
\ldots,\k(\gamma_{s,m-1})
\label{differentOrdering}
\eeq
is the natural choice. Since the matrix $\Lambda=F_s^\dagger GF_s$ is
positive definite as $G$, also $D=\Pi\Lambda\Pi^T$ and its diagonal
blocks $D_k$ are positive definite. Then
\beq
D^{1/2}=\Pi\Lambda^{1/2}\Pi^T=\left[\begin{matrix}D_1^{1/2}&\ldots&0\cr
\vdots&\ddots&\vdots\cr0&\ldots&D_s^{1/2}\cr\end{matrix}\right]
\eeq
is well defined. Applying to $D^{1/2}$ the inverse permutation $\Pi^T$
one gets $G^{1/2}=F_s\Lambda^{1/2}F_s^{-1}$, where
\beq
\Lambda^{1/2}=\Sigma=\left[\begin{matrix}\Sigma_{11}&\ldots&\Sigma_{1s}\cr
							\vdots&\ddots&\vdots\cr
							\Sigma_{s1}&\ldots&\Sigma_{ss}\cr\end{matrix}\right]\;.
\eeq
where the blocks $\Sigma_{hk}$ are diagonal matrices of order $m$.

The diagonal blocks of the square root of the weighted Gram matrix
are $(G^{1/2})_{hh}=F\Sigma_{hh}F^\dagger$. This results in a
circulant matrix with diagonal elements, say $g_h$, depending on
$h$. The condition of optimality of the SRM (see Theorem 3) is
satisfied if and only if $g_h=g$ is independent of $h$. This is
summarized by the following theorem.

\mn{\sc Theorem 4.} The SRM is optimal for the multiple constellation
of GUS states if and only if the diagonal blocks of the square root
$G^{1/2}$ of the weighted Gram matrix have equal diagonal entries.
$\hfill\square$

\medskip

An alternative criterion derives from the fact that
\beq
g_h=\frac{1}{m}\Tr((G^{1/2})_{hh})=\frac{1}{m}\Tr(F\Sigma_{hh}
F^\dagger)=\frac{1}{m}\Tr(\Sigma_{hh})\;.
\label{traceCriterion}
\eeq
Then the condition of optimality of the SRM  is satisfied if and
only if the blocks $\Sigma_{hh}$ have equal traces.

Finally, provided that the SRM is optimal, the correct decision
probability becomes
\beq
P_c=msg^2\;.
\label{defPcSRM}
\eeq

\section{Double quantum PSK constellation}

As a first application of the above theory we consider a double quantum
{\it binary phase shift keying} (BPSK) with four coherent states $\k(\pm
\alpha)$ and $\k(\pm\beta)$ in a Fock space. The system is a particular case
of the scheme discussed in the previous section by setting
\beq
\k(\gamma_{10})=\k(\alpha),\ \k(\gamma_{11})=\k(-\alpha),\ 
\k(\gamma_{20})=\k(\beta),\ \k(\gamma_{21})=\k(-\beta)\;.
\label{DoublePSKStates}
\eeq
In this case the symmetry operator is the rotation operator $S=e^{i\pi
a^\dagger a}$ with $a$ and $a^\dagger$ annihilation and creation operators
in the Fock space giving $S\k(\alpha)=\k(-\alpha)$. According to our
previous assumption the probabilities of $\k(\pm\alpha)$ and $\k(\pm\beta)$
are $p$ and $q$ respectively, with $p+q=1/2$.

The Gram matrix depends on the inner products 
\beq
\chi=\bk(\alpha,\beta),\ 
	\eta_\alpha=\bk(\alpha,-\alpha),\ \eta_\beta
	=\bk(\beta,-\beta),\ \xi=\bk(\alpha,-\beta)\;.
\eeq
The blocks of the weighted Gram matrix turn out to be
\begin{align}
& G_{11}=p\left[\begin{matrix}1&\eta_\alpha\cr \eta_\alpha&1\cr\end{matrix}\right],
 \quad G_{22}=q\left[\begin{matrix}1&\eta_\beta\cr \eta_\beta&
1\cr\end{matrix}\right], \\
& \qquad G_{12}=\sqrt{pq}\left[\begin{matrix}\chi&\xi\cr \xi&
\chi\cr\end{matrix}\right]=G_{21}^*.
\end{align}
Applying the discrete Fourier transform \eqref{B5} gives the eigenvalues of the block $G_{11}$
\beq
\lambda_0 = p(1+\eta_\alpha), \qquad \lambda_1 = p(1-\eta_\alpha).
\eeq
Similarly, the eigenvalues of $G_{12}$ are
\beq
\mu_0 = \sqrt{pq}(\chi+\xi), \qquad \mu_1 = \sqrt{pq}(\chi-\xi),
\eeq
and the eigenvalues of $G_{22}$
\beq
\omega_0 = p(1+\eta_\beta), \qquad \omega_1 = p(1-\eta_\beta).
\eeq

The weighted Gram matrix is decomposed as $G=F_2 \Lambda F_2^{\dagger}$, where $\Lambda$ is the matrix \eqref{defLambda} with diagonal blocks
$\Lambda_{11} = \diag\{\lambda_0, \lambda_1\}$, $\Lambda_{12}=\diag\{\mu_0, \mu_1\}$ and $\Lambda_{22}=\diag\{\omega_0, \omega_1\}$.

The square root $\Sigma=\Lambda^{1/2}$ can be evaluated taking advantage of the sparsity of the matrix $\Lambda$, or, as an alternative, considering the permutation $\Pi$ that converts the ordering of the states into \eqref{differentOrdering}. The result is the block-diagonal matrix  \eqref{defD}, which is related to $\Lambda$ through $D=\Pi \Lambda \Pi^T$, and has diagonal blocks
\begin{align}
& D_{1}=\left[\begin{matrix}p(1+\eta_\alpha)&\sqrt{pq}(\chi+\xi)\cr \sqrt{pq}(\chi+\xi)^*&q(1+\eta_\beta)\end{matrix}\right], \\
& D_{2}=\left[\begin{matrix}p(1-\eta_\alpha)&\sqrt{pq}(\chi-\xi)\cr \sqrt{pq}(\chi-\xi)^*&q(1-\eta_\beta)\end{matrix}\right].
\end{align}

The square root $D^{1/2}$ has the same structure of $D$, and can be evaluated from the square root of $D_{1}^{1/2}$ and $D_{2}^{1/2}$, with
\begin{align}
&D_{1}^{1/2}=\frac{1}{\sqrt{p(1+\eta_\alpha)+q(1+\eta_\beta)+2\sqrt{\Delta^{(+)}}}} \nonumber \\
& \qquad \times \left[\begin{matrix}p(1+\eta_\alpha)+\sqrt{\Delta^{(+)}}&\sqrt{pq}(\chi+\xi)\cr \sqrt{pq}(\chi+\xi)^*&q(1+\eta_\beta)+\sqrt{\Delta^{(+)}}\end{matrix}\right],
\label{SquareRootD}
\end{align}
where $\Delta^{(+)}=pq[(1+\eta_\alpha)(1+\eta_\beta)-|\chi+\xi|^2]$. Similarly the square root of $D_2$, which can also be obtained from \eqref{SquareRootD} with the substitutions 
\beq
\eta_\alpha \to -\eta_\alpha,\quad \eta_\beta \to -\eta_\beta,\quad \xi \to - \xi.
\label{SquareRootSubs}
\eeq

Finally, the square root $\Sigma=\Lambda^{1/2}$ is obtained applying the inverse permutation $\Pi^T$ to the block-diagonal matrix $D^{1/2}$. The blocks of $\Sigma$ are still diagonal, and in particular
\begin{align}
& \Sigma_{11}=\diag \left  \{\frac{p(1+\eta_\alpha)+\sqrt{\Delta^{(+)}}}{\sqrt{p(1+\eta_\alpha)+q(1+\eta_\beta)+2\sqrt{\Delta^{(+)}}}} \ , \right. \nonumber \\
& \qquad \qquad \quad \left. \frac{p(1-\eta_\alpha)+\sqrt{\Delta^{(-)}}}{\sqrt{p(1-\eta_\alpha)+q(1-\eta_\beta)+2\sqrt{\Delta^{(-)}}}} \right \} \label{BlockSigma11}
\end{align}
and
\begin{align}
& \Sigma_{22}= \diag \left \{ \frac{q(1+\eta_\beta)+\sqrt{\Delta^{(+)}}}{\sqrt{p(1+\eta_\alpha)+q(1+\eta_\beta)+2\sqrt{\Delta^{(+)}}}} \ , \right. \nonumber \\
& \qquad \qquad \quad \left. \frac{q(1-\eta_\beta)+\sqrt{\Delta^{(-)}}}{\sqrt{p(1-\eta_\alpha)+q(1-\eta_\beta)+2\sqrt{\Delta^{(-)}}}}\right \} \label{BlockSigma22}
\end{align}
with $\Delta^{(-)}$ obtained from $\Delta^{(+)}$ with the substitutions \eqref{SquareRootSubs}. The square root of the weighted Gram matrix is then obtained through the Fourier transform, $G^{1/2}=F_2\Lambda^{1/2}F_2^{-1}$.

The optimality of SRM can be verified with the aid of Theorem 4, or, as an alternative, through the conditions \eqref{traceCriterion}. Given a set of parameters $\alpha, \ \beta$ and $p$, the optimality is not assured. However, it is worthwhile to investigate how one of the parameters should be optimized given the others, in order to meet the optimality conditions. This is the topic of this section, where we investigate the optimization of the variable $p$ given the parameters $\alpha,\ \beta$. Although this optimization can be (numerically) performed for arbitrary $\alpha,\ \beta$, we focus our attention on two particular cases.

\subsection{Case $|\alpha|=|\beta|$}

As a first case, we consider $|\alpha|=|\beta|$. Practical application of this 
case arises
when the transmitter, supposed to produce a {\sf Quaternary-PSK}, has a misalignment or a systematic bias error in the angle defining one of the two constellations. Without loss of generality, the alphabet can be defined with two real parameters, $\alpha \in \mathbb{R}$ and $\beta=\alpha e^{i \delta}$, with  $\delta \in [0,\pi/2]$ defining the angular shift between the two binary constellations. 

The entries in the weighted Gram matrix $G$ simplifies, and the inner products between the states become 
$$
\chi=e^{-\alpha^2(1-e^{i\delta})},\ 
\eta_\alpha=\eta_\beta=e^{-2\alpha^2},\ 
\xi=e^{-\alpha^2(1+e^{i\delta})}.
$$
It is immediate to verify that with these positions, condition \eqref{traceCriterion} for the optimality of the SRM is satisfied by $p=q=1/4$. That is, the equal prior probability between the symbols gives the optimality of SRM, even in the case $\delta \neq \frac{\pi}{2}$.

The probability of correct detection is obtained employing \eqref{traceCriterion} and \eqref{defPcSRM}, resulting in
\begin{align*}
&P_c=\frac{1}{16}\left(\sqrt{1+\eta_{\alpha}+|\chi+\xi|} + \sqrt{1+\eta_{\alpha}-|\chi+\xi|} \right. \nonumber \\
& \qquad \qquad \left. + \sqrt{1-\eta_{\alpha}+|\chi-\xi|} + \sqrt{1-\eta_{\alpha}-|\chi-\xi|}\right)^2
\end{align*}

\begin{figure}
\centering
	\includegraphics[width=0.45\textwidth]{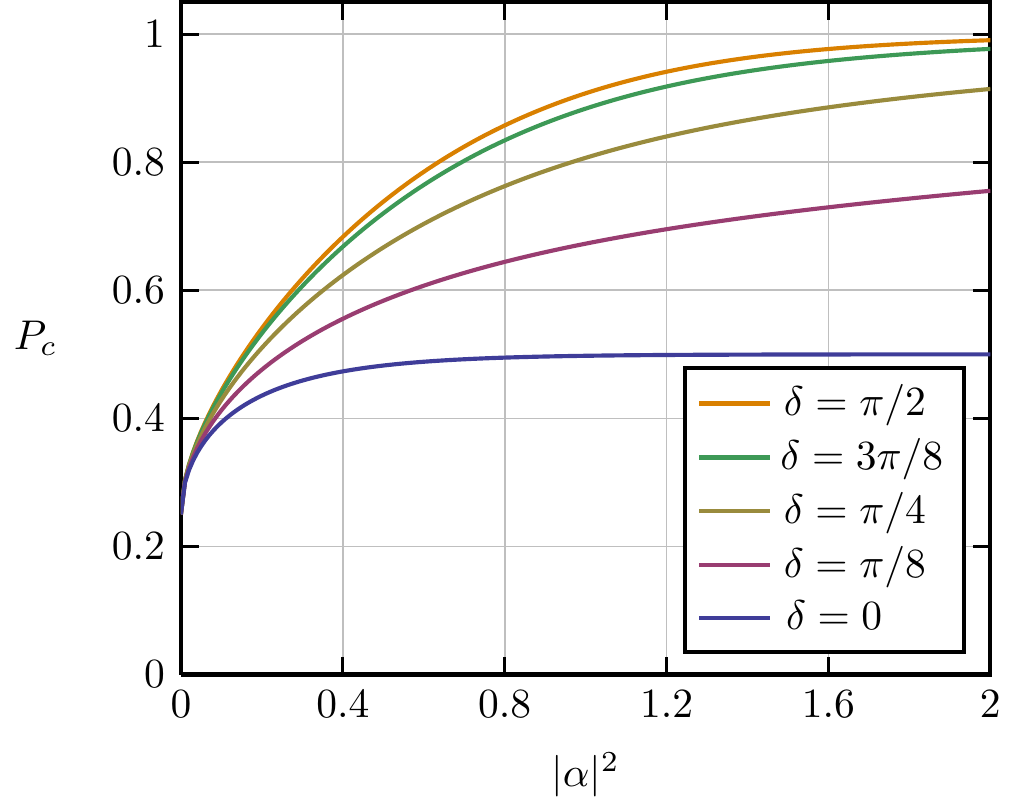}
	\caption{(Color online) Probability of correct detection for a Double BPSK constellation, as a function of the mean photon number $|\alpha|^2$ employed in the transmission of each state. From the top to the bottom, the performance of different values of $\delta$ as they appear in the legend.}
	\label{Pc_DoublePSK}
\end{figure}

In Fig. \ref{Pc_DoublePSK} the performances of the Double BPSK are plotted as a function of the mean photon number $|\alpha|^2$ employed in each transmitted state. In the figure, different curves correspond to the values of the phase shift $\delta=0,\ \pi/8,\ \pi/4,\ 3\pi/8, \pi/2$. As we can see, the lines are monotonically increasing in both the value of $|\alpha|^2$ and $\delta$. In addition, the performances are similar when $\delta$ is around $\pi/2$, which is the situation of the Quaternary PSK, while they drop quickly for lower values of the phase shift.

\subsection{Case $\beta=3\alpha$}

As a second case of study, we consider $\alpha \in \mathbb{R},\ \beta = 3 \alpha$. This set of constellations corresponds to the modulation scheme known as {\sf 4-Pulse Amplitude Modulation} (PAM). While usually an equal prior probability is assumed for the transmitted states, here we study the optimization of this probability such that the SRM are optimal.

The inner products between the states read
\begin{align*}
& \chi=\bk(\alpha,3\alpha)=e^{-2\alpha^2} = \eta_\alpha,\\
& \eta_\beta=\bk(3\alpha,-3\alpha)=e^{-18\alpha^2}=\eta_\alpha^9,\\
& \xi=\bk(\alpha,-3\alpha)=e^{-8\alpha^2}=\eta_\alpha^4.
\end{align*}
Again, we resort to conditions \eqref{traceCriterion} in order to find the value of $p$ that makes the SRM optimal. In this case, there are no evident solutions, and the authors could not find any closed form solution, so we turn to numerical algorithms to find the optimal value. 

\begin{figure}
\centering
	\includegraphics[width=0.45\textwidth]{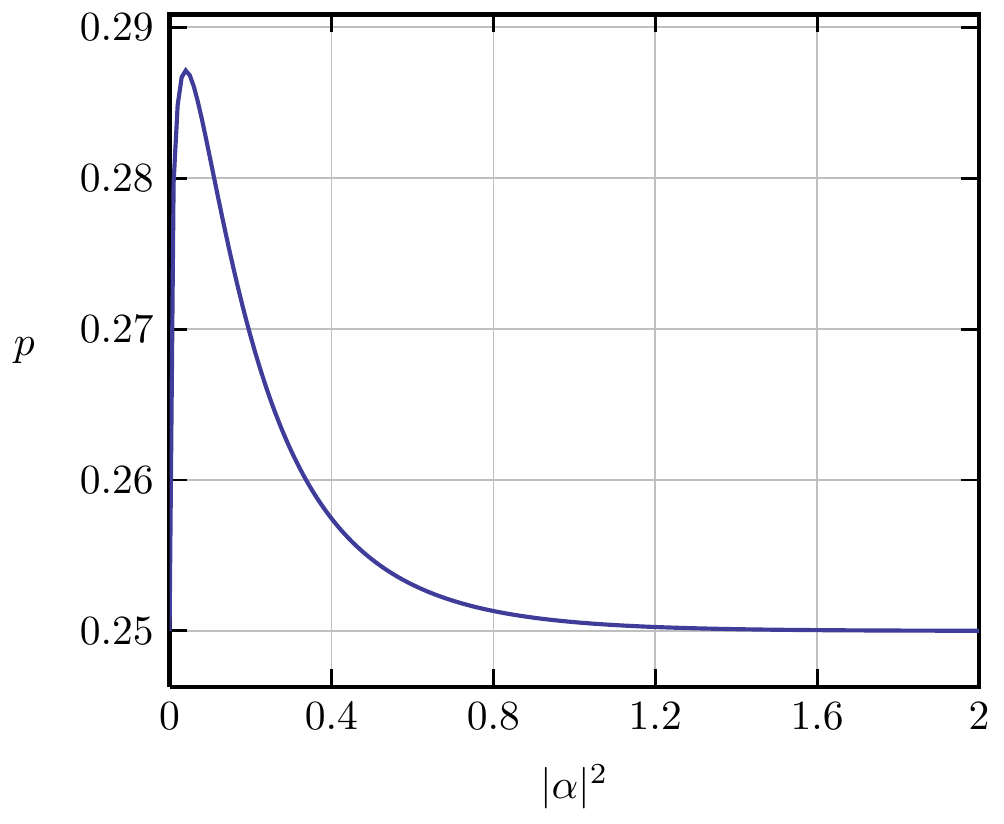}
	\caption{(Color online) Optimal prior probability $p$ of the state $\ket{\pm\alpha}$ of the Double BPSK modulation, in the case $\alpha \in \mathbb{R}$, $\beta~=~3\alpha$. 
	}
	\label{optPriori4PAM}
\end{figure}

\begin{figure}
\centering
	\includegraphics[width=0.45\textwidth]{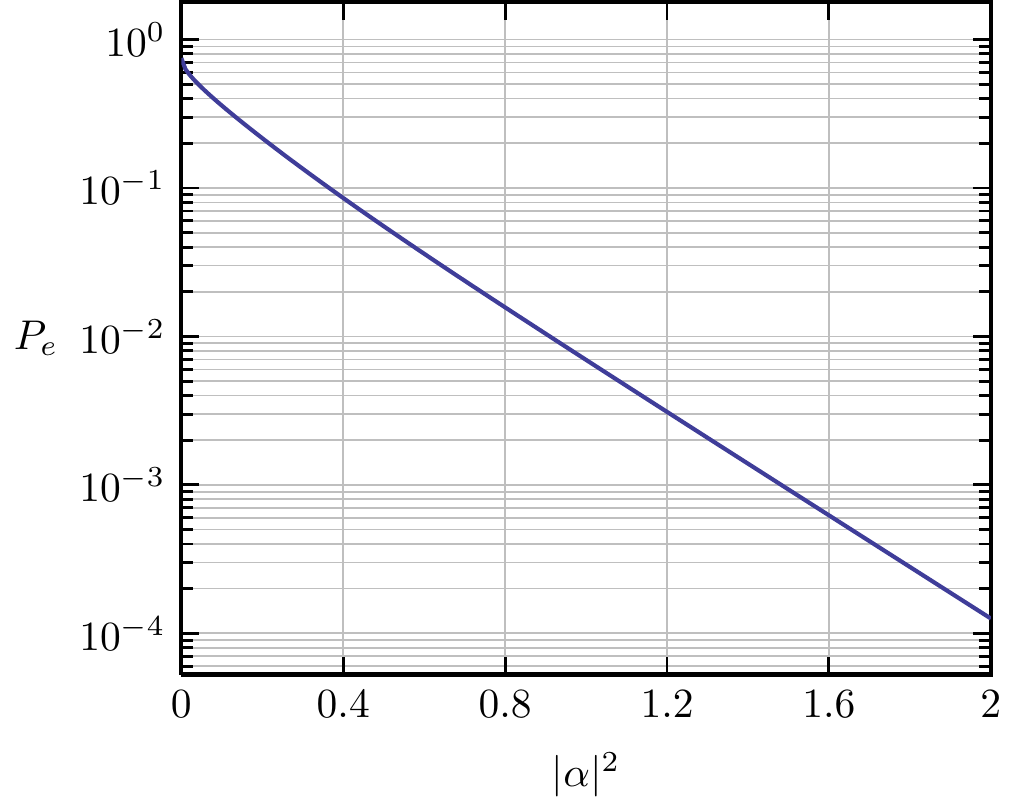}
	\caption{(Color online) Probability of error of a 4-PAM system with optimized a priori probability and Square Root Measurement at the receiver. On the $x$ axis, the mean photon number employed in the transmitted state $\ket{\pm \alpha}$.}
	\label{Pe4PAM}
\end{figure}

The results of the numerical optimization are plotted in Fig. \ref{optPriori4PAM}, as a function of the mean photon number $|\alpha|^2$ of the states $\ket{\pm\alpha}$. Note that while for low values of $\alpha$ the prior $p$ moves away from equal distribution, the optimal solution tends to $p=0.25$ as the value of $\alpha$ increases. The corresponding performance is plotted in Fig. \ref{Pe4PAM}, which show an increase in the performance as the value of $|\alpha|^2$ increases.

We have compared the performance of the SRM with optimal a priori probabilities
with the performances of both the SRM and the optimal measurement (evaluated 
via semidefinite programming \cite{Eldar2003}) with equal a priori probabilities. The results
are practically indistinguishible if plotted as in Fig. \ref{Pe4PAM}. This further comfirms that 
SRM is really a very good measurement in cases of practical interest. In particular,
in the present case, optimizing the a priori probabilities does not reduces in 
significant way the error probability.



\section{Double quantum PPM constellation}

\subsection{Simple quantum PPM constellation}

A popular scheme proposed for deep space quantum communications is
the {\it pulse position modulation} (PPM) \cite{DescansoCap4}. This scheme can be
modeled in a Hilbert space $\C(H)^{\otimes m}$ formed by $m$
replicas of the Fock space $\C(H)$. The states used by Alice are $m$
tensor products
\beq
\k(\gamma_i)=\k(\gamma_{i0})\otimes\ldots\otimes\k(\gamma_{i,m-1})
\eeq
where only $\k(\gamma_{ii})$ coincides with a coherent state $\k(
\alpha)$, while the other coincide with the null state $\k(0)$. For
instance, for $m=3$, the states are
\begin{align}
& \k(\gamma_0)=\k(\alpha)\otimes\k(0)\otimes\k(0), \nonumber \\
& \k(\gamma_1)=\k(0)\otimes\k(\alpha)\otimes\k(0), \\
& \k(\gamma_2)=\k(0)\otimes\k(0)\otimes\k(\alpha). \nonumber 
\end{align}
In practical realizations, in correspondence with the symbol $i$ the
laser radiates only in the $i$--th slot of a time frame of $m$ slots.

Although the symmetry of the states is apparent (every state is a cyclic
permutation of the preceding), the evaluation of the unitary symmetry
operator $S$ in the tensor space $\C(H)^{\otimes m}$ is by no means
trivial \cite{Cariolaro2010}. In any case, provided that the states have
equal probabilities $1/m$, the weighted Gram matrix is given by
\beq
G=\frac{1}{m}\left[\begin{matrix}1&\chi&\ldots&\chi\cr
\chi&1&\ldots&\chi\cr\vdots&\vdots&\ddots&\vdots\cr\chi&\chi&\ldots&1
\cr\end{matrix}\right]
\eeq
with $\chi=\bk(\alpha,0)\bk(0,\alpha)=e^{-\alpha^2}$ (without loss of
generality $\alpha$ is assumed real). Since $G^{1/2}$ is circulant as $G$,
the SRM is optimal. Applying the discrete Fourier transform (B5) gives the
eigenvalues of $G$
\begin{align}
&\lb_0=\frac{1}{m}[\,1+(m-1)\chi\,],  \\
&\lb_k=\frac{1}{m}(1-\chi),\qquad   k=1,\ldots m-1\;.\nonumber
\end{align}
Applying to the eigenvalues $\lb_k^{1/2}$ of $G^{1/2}$ the inverse discrete
Fourier transform (B6), one gets the first row of $G^{1/2}$, namely,
\begin{align*}
& c_0=\frac{1}{m\sqrt m}\left[\sqrt{1+(m-1)\chi}+(m-1)\sqrt{1-\chi}\right], \\
& c_1=\ldots=c_{m-1}=\frac{1}{m\sqrt m
}\left[\sqrt{1+(m-1)\chi}-\sqrt{1-\chi}\right].
\end{align*}
Note that $G^{1/2}$ has the same structure of $G$ with equal entries out
of the diagonal.

The correct detection probability is obtained by \eqref{defPcWithLambda} and is
$$
P_c=mc_0^2=\frac{1}{m^2}\left[\sqrt{1+(m-1)\chi}+(m-1)
	\sqrt{1-\chi}\right]^2
$$

\subsection{Double quantum PPM constellation}

Here we propose a new PPM scheme with the goal of doubling the
number of the states and, possibly, of improving the capacity of the
quantum communication system. We consider a double constellation
in $\C(H)^{\otimes n}$
$$
\k(\gamma_{10}),\ldots,\k(\gamma_{1,m-1})\qquad\k(\gamma_{20}),\ldots,
\k(\gamma_{2,m-1})\;.
$$
where, as in the ordinary quantum PPM, $\k(\gamma_{1i})$ includes only
one non null state $\k(\alpha)$ in the $i$--th slot, while $\k(\gamma_{2i})$
includes only one non null state $\k(-\alpha)$ in the $i$--th slot. To
maintain the symmetry of the problem, we assume that the states have
equal probability $1/(2m)$. The weighted Gram matrix turns out to to be
$$
G=\left[\begin{matrix}H&K\cr K&H\cr\end{matrix}\right]
$$
where $H_{ij}=\bk(\gamma_{0i},\gamma_{0j})=\bk(\gamma_{1i},
\gamma_{1j})$ and $K_{ij}=\bk(\gamma_{0i},\gamma_{1j})=
\bk(\gamma_{1i},\gamma_{0j})$. Simple considerations lead to the
following circulant matrices
$$
H=\frac{1}{2m}\left[\begin{matrix}1&\chi&\ldots&\chi\cr
\chi&1&\ldots&\chi\cr\vdots&\vdots&\ddots&\vdots\cr\chi&\chi&\ldots&1
\cr\end{matrix}\right], \quad 
K=\frac{1}{2m}\left[\begin{matrix}\chi^2&\chi&\ldots&\chi\cr
\chi&\chi^2&\ldots&\chi\cr\vdots&\vdots&\ddots&\vdots\cr\chi&\chi
&\ldots&\chi^2\cr\end{matrix}\right].
$$
The discrete Fourier transform \eqref{B5} gives the eigenvalues of $H$
\begin{align}
& \lb_0=\frac{1}{2m}[\,1+(m-1)\chi\,], \\
& \lb_k=\frac{1}{2m}(1-\chi),\quad k=1,\ldots,m-1\;. \nonumber
\end{align}
and the eigenvalues of $K$
\begin{align}
& \mu_0=\frac{1}{2m}[\,\chi^2+(m-1)\chi\,], \\
& \mu_k=\frac{1}{2m}(\chi^2-\chi),\quad k=1,\ldots,m-1\;. \nonumber
\end{align}
Then $G=F_2\Lambda F_2^\dagger$ with
\beq
\Lambda=\left[\begin{matrix}\Lambda_0&\Lambda_1\cr \Lambda_1&
\Lambda_0\cr\end{matrix}\right]
\eeq
where $\Lambda_0=\diag\{\lb_0,\ldots,\lb_{m-1}\}$ and
$\Lambda_1=\diag\{\mu_0,\ldots,\mu_{m-1}\}$. It follows that
\beq
G^{1/2}=F_2\left[\begin{matrix}\Sigma_0&\Sigma_1\cr
\Sigma_1&\Sigma_0\cr\end{matrix}\right]F_2^\dagger=\left[\begin{matrix}R&T\cr
T&R\cr\end{matrix}\right] \;,
\eeq
where the diagonal matrices $\Sigma_0$ and $\Sigma_1$ satisfy the
conditions $\Sigma_0^2+\Sigma_1^2=\Lambda_0$ and
$2\Sigma_0\Sigma_1=\Lambda_1$. Since the diagonal blocks of
$\Sigma$ coincide, the SRM  is optimal. Simple computations lead to
$\Sigma_{0}=\diag\{\nu_0,\nu_1,\ldots,\nu_1\}$ and $\Sigma_{1}=
\diag\{\xi_0,\xi_1,\ldots,$ $\xi_{m-1}\}$ with
\begin{align*}
& \nu_0=\frac{1}{\sqrt{8m}}\left[\sqrt{1+\chi^2+2(m-1)\chi}
	+\sqrt{1-\chi^2}\right], \\
& \nu_1=\frac{1}{\sqrt{8m}}\left[1-\chi+\sqrt{1-\chi^2}\right], \\
& \xi_0=\frac{1}{\sqrt{8m}}\left[\sqrt{1+\chi^2+2(m-1)\chi}
	-\sqrt{1-\chi^2}\right], \\
& \xi_1=\frac{1}{\sqrt{8m}}\left[1-\chi-\sqrt{1-\chi^2}\right].
\end{align*}
Finally, using the inverse discrete transform \eqref{B6}, we get the first rows
of the circulant matrices $R$ and $T$
\begin{align*}
& r_0=\frac{1}{2m\sqrt{2m}}\left[
\sqrt{1+2(m-1)\chi+\chi^2} \right. \\
& \qquad \qquad \qquad \qquad \left. +(m-1)(1-\chi)+m\sqrt{1-\chi^2}\right] \\
& t_0=\frac{1}{2m\sqrt{2m}}\left[
\sqrt{1+2(m-1)\chi+\chi^2} \right. \\
& \qquad \qquad \qquad \qquad \left. +(m-1)(1-\chi)-m\sqrt{1-\chi^2}\right] \\
& r_i=t_i=\frac{1}{2m\sqrt{2m}}\left[\sqrt{1+2(m-1)\chi+\chi^2}-(1-\chi)\right], \\
& \qquad \qquad \qquad \qquad \qquad \qquad \qquad \qquad i=1,\ldots,m-1\;.
\end{align*}
Note that $R$ and $T$ have equal entries out of the diagonal as a
consequence of the particular symmetry.

Finally \eqref{defPcSRM} gives the correct decision probability of the double PPM
\begin{align*}
& P_c=2mr_0^2=\frac{1}{4m^2}\left [\sqrt{1+\chi^2+2(m-1)\chi} \right. \\
& \qquad \qquad  \qquad \left. +(m-1)(1-\chi)+m\sqrt{1-\chi^2}\right]^2
\end{align*}

\subsection{Mutual information comparison}

\begin{figure}
\centering
	\includegraphics[width=0.45\textwidth]{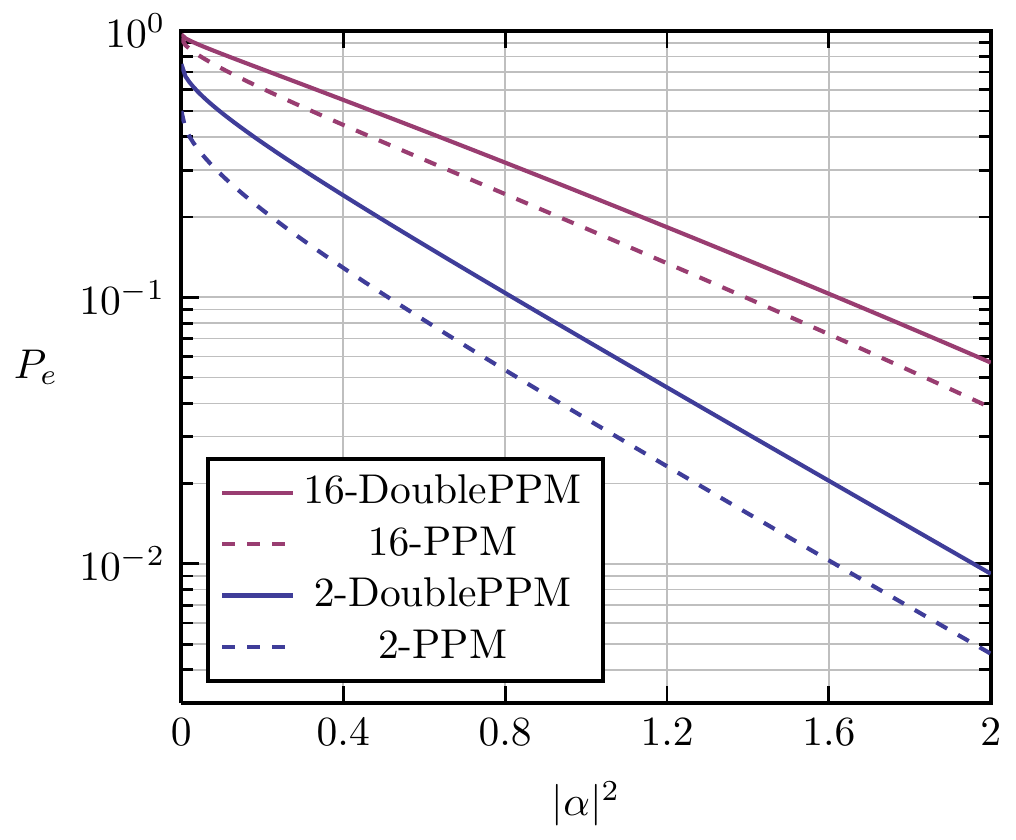}
	\caption{(Color online) Probability of error of $m$-PPM (dashed) and $m$-Double PPM (solid), for $m=2$ (lower lines) and $m=16$ (upper lines). On the $x$ axis, the mean photon number employed in the transmitted state, $|\alpha|^2$.}
	\label{Pe_PPMvsDoppioPPM}
\end{figure}

In Fig. \ref{Pe_PPMvsDoppioPPM} the error probabilities of the simple and double PPM are
compared for $m=2$ and $m=16$ in terms of the mean number of photons
per symbol $\alpha^2$ related to the parameter $\chi$ by $\chi=e^{
-\alpha^2}$. The error probability of double PPM is larger, but its states
are twice as many. Then a significant comparison requires the evaluation
of the informations transferred by the systems.

The channel defined by the simple PPM is a symmetric channel with joint
input--output probabilities $p\,(i,i)=c_0^2$ and $p\,(i,j)=c_1^2$ for $j\neq i$.
The marginal probabilities are uniform $p\,(i)=p\,(j)=1/m$. Then the mutual
information turns out to be
\begin{align}
I_1& =\sum_{i,j=0}^{m-1}p\,(i,j)\log{\frac{p\,(i,j)}{p\,(i)p\,(j)}} \\
& = 2\log m	+ mc_0^2\log c_0^2+m(m-1)c_1^2\log c_1^2  \nonumber
\end{align}

The channel defined by the double PPM has joint input--output
probabilities $p\,(i,i)=c_0^2$, $p\,(i,m+i)=p\,(m+i,i)=t_0^2$, $i=0,\ldots,m-1
$, while the other $4m^2-4m$ crossover probabilities have  common value
$r_i$. The marginal probabilities are uniform $p\,(i)=p\,(j)=1/(2m)$ as a
consequence of the symmetry. In conclusion the mutual information turns
out to be
\begin{align}
I_2 & =\sum_{i,j=0}^{2m-1}p\,(i,j)\log{\frac{p\,(i,j)}{
p\,(i)p\,(j)}} =2\log(2m)+2mr_0^2\log r_0^2  \nonumber \\
& \qquad + 2mt_0^2\log t_0^2+4(m-1)mr_i^2\log r_i^2
\end{align}

\begin{figure}
\centering
	\includegraphics[width=0.45\textwidth]{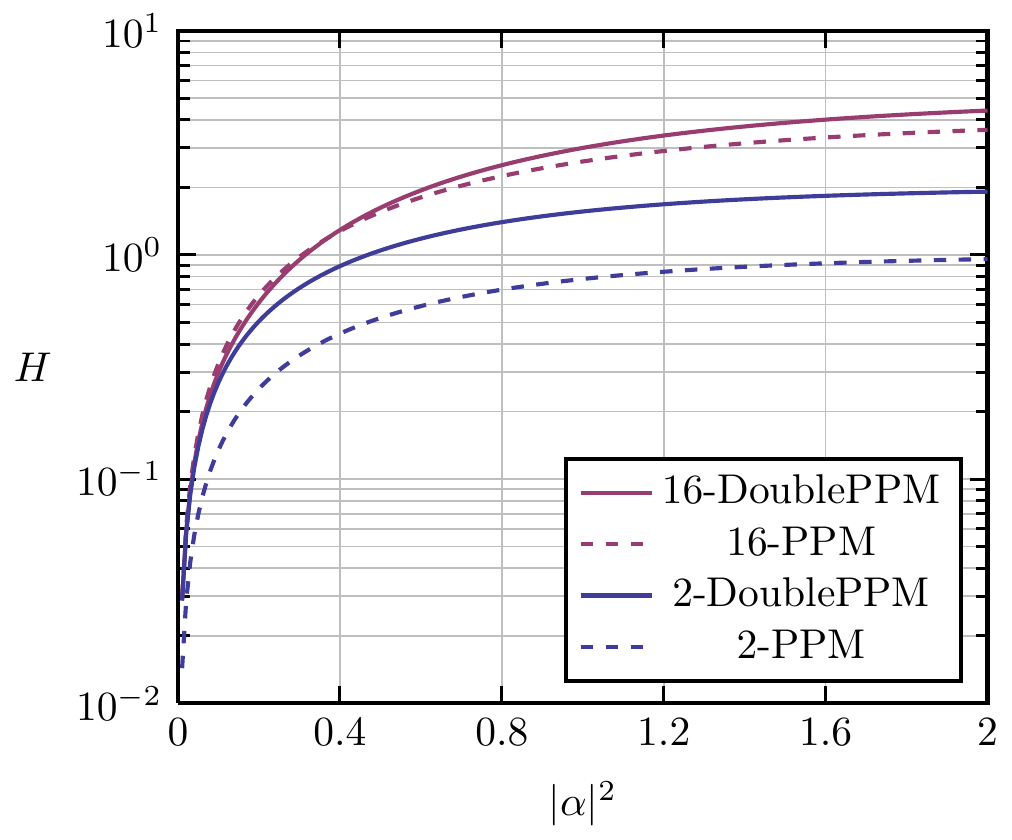}
	\caption{(Color online) Mutual information between the transmitted and estimated symbols, for a $m$-PPM (dashed) and $m$-Double PPM (solid) constellation, in the case of $m=2$ (lower lines) and $m=16$ (upper lines). On the $x$ axis, the mean photon number employed in the transmitted state, $|\alpha|^2$.}
	\label{H_PPMvsDoppioPPM}
\end{figure}

Asymptotically, as $\alpha$ increases (and $\chi$ tends to zero), $I_1$ tends
to $\log m$ and $I_2$ tends to $\log(2m)$ with a gain of one bit per symbol.
The mutual informations $I_1$ and $I_2$, coinciding with  the capacities
of the channels, are compared in Fig. \ref{H_PPMvsDoppioPPM} for $m=2$ and $m=16$.

Two remarks are adequate. First, the double PPM scheme is a combination
of the simple PPM scheme and of a binary PSK scheme. Further gain
could be obtained combining PPM with $k$--PSK, using $k$ PPM
constellation with non empty states $\k(\alpha e^{i2\pi r/k})$, $r=0,1,\ldots,
k-1$. Second, from a practical point of view, double PPM implies some
complications, both on the transmitter and on receiver side. At the
transmitter the on--off modulation must be replaced by a phase modulation,
while at the receiver a simple photon counter must be replaced by a phase
sensitive device.  We do not insist on these topics that are beyond the scope
of the paper.

\section{Conclusions}

The square root measurement furnishes an alternative approach to the
reliable discrimination of quantum states. This measurement is only
suboptimal, in general, but it is well known that it turns to be optimal for
quantum states enjoying geometrically uniform simmetry. In this paper, we
showed that the square root measurement is optimal also in other situations
of practical interest. In particular we found necessary and sufficient
conditions for the optimality in the presence of multiple constellations
of symmetrical states also with non uniform probabilities. As an
application example we considered the case of two pairs of quantum binary
symmetrical states (PSK states). Finally, the theory is applied to a possible
improvement of the pulse position modulation (PPM) scheme that increases
the capacity of the resulting channel.

\acknowledgments

Nicola Dalla Pozza acknowledges partial support by the `Borsa Gini' scholarship, awarded by `Fondazione Aldo Gini', Padova, Italy.

\appendix

\section{Proof of Theorem 2}

We apply the Theorem 1 to the operators $W_i=\kb(\tgamma_i,\tgamma_i)$
and $P_i=\kb(\mu_i,\mu_i)$ and preliminarily note the following matrix
representations of $Y$, $W_r$ and $Y^{(r)}=Y-W_r$ in terms of the
orthonormal basis $\{\k(\mu_i)\}$:
\begin{align}
 &\bok(\mu_i,Y,\mu_j)=X_{ij}X_{jj}^*\;,\quad \bok(\mu_i,W_r,\mu_j)=X_{ir}X_{jr}^*\;, \nonumber \\
& Y^{(r)}_{ij}=X_{ij}X_{jj}^*-X_{ir}X_{jr}^*\;.
\label{A1}
\end{align}
If the measurement is optimal statement i) follows directly from \eqref{condizioniOttimeUguaglianzaZero} and
$Y=XX_d^*$ is semidefinite positive. Since $X$ is not singular, it
remains to prove that the diagonal entries of $X_d$ (of $X$) are different
from 0. Assume that $X_{00}=\bk(\mu_0,\tilde\gamma_0)=0$. Then, it
must exist $j\neq 0$ such that $X_j=\bk(\mu_0,\tilde\gamma_j)\neq 0$,
for otherwise all the states $\k(\tilde\gamma_j)$ would be orthogonal to
$\k(\mu_0)$ against the assumption of their linear independence.
Without loss of generality, assume that $X_{01}\neq0$. As a
consequence of i) we have also $X_{11}=0$. Now, if we change the roles
of the measurement vectors $\k(\mu_0)$ and $\k(\mu_1)$, the matrix $X$ is
modified in $\widehat X$, with $\widehat X_{jj}=X_{jj}$ for $j>1$,
$\widehat X_{00}=X_{10}$, and $\widehat X_{11}=X_{01}$. Denoted
by $\widehat P_c$ the new correct decision probability, we get
$\widehat P_c-P_c=|X_{01}|^2+|X_{10}|^2>0$ contradicting the optimality
of  $X$. One concludes that $X_d$ has non zero diagonal entries and
$XX_d^*$ is definite positive. In order to prove the sufficiency note that
the matrices $Y^{(r)}$  are Hermitian and $Y=Y^{(r)}+\kb(\tgamma_r,
\tgamma_r)$. Then they satisfy the conditions of a theorem on the
eigenvalues of Hermitian matrices \cite[Theorem 4.3.4]{HornJohnson} which states
that, denoted  by $\nu_i$ the eigenvalues of $Y^{(r)}$ and by $\lb_i$ the
eigenvalues of $Y$ arranged in increasing order, we have
\beq
\nu_1\le\lb_1\le\nu_2\le\ldots\le\nu_m\le\lb_m\;.
\label{A2}
\eeq
If $Y$ is positive definite, it follows that $\lb_1>0$. Since in $Y^{(r)}$ the
$r$--th column vanishes, at least one of its eigenvalues vanishes and $\nu_1
=0$. One concludes that all eigenvalue of $Y^{(r)}$ are non negative and
$Y^{(r)}$ is semidefinite positive. Since this holds true for each $r$, the
condition \eqref{condizioniOttimeUguaglianzaZero} of Theorem 1 is satisfied and the operators $P_i=\kb(\mu_i,
\mu_i)$ provide the optimal measurement.

\section{Circulant matrices}

We collect in this appendix for convenience some properties of the
circulant matrices which are used in the paper. For more details the
reader is deferred to the literature on the topic, for instance
\cite{Davis1979}.

A matrix $G$ of order $m$ is said {\it circulant} if
\beq
G=\left[\begin{matrix}c_0&c_1&\ldots&c_{m-1}\cr
c_{m-1}&c_0&\ldots&c_{m-2}\cr
\vdots&\vdots&\ddots&\vdots\cr
c_1&c_2&\ldots&c_0\cr\end{matrix}\right]\;,
\label{B1}
\eeq
i.e., if
\beq
G_{ij}=c_{(j-i){\rm mod}\, m}\;.
\label{B2}
\eeq
In other words $G$ is circulant if its rows are cyclic permutation
of the first row. A matrix $G$ is circulant if and only if has
spectral decomposition
\beq
G=F\Lambda F^\dagger
\label{B3}
\eeq
where $F$ is the (unitary) {\it  Fourier matrix} with entries
\beq
F_{hk}=\frac{1}{\sqrt{m}}e^{i2\pi kh/m},\qquad h,k=0,
	\ldots,m-1
\label{B4}
\eeq
and the diagonal matrix $\Lambda={\rm diag}\{\lb_0,\ldots,\lb_{m-1}\}$
collects the eigenvalues of $G$. The circulant matrices of order $m$
form a multiplicative commutative group of matrices simultaneously
diagonalizable. The eigenvalues of the circulant matrix $G$ can be
obtained from the first row of the matrix via the {\it discrete
Fourier transform}
\beq
\lb_k=\sum_{r=0}^{m-1}c_re^{i2\pi kr/m}\;.
\label{B5}
\eeq
Finally, the inverse discrete Fourier transform gives the first row
of $G$
\beq
c_r=\frac{1}{m}\sum_{k=0}^{m-1}\lb_ke^{-i2\pi kr/m}\;.
\label{B6}
\eeq
\bibliographystyle{apsrev}
\bibliography{SRMbiblio}

\begin{thebibliography}{26}
\expandafter\ifx\csname natexlab\endcsname\relax\def\natexlab#1{#1}\fi
\expandafter\ifx\csname bibnamefont\endcsname\relax
  \def\bibnamefont#1{#1}\fi
\expandafter\ifx\csname bibfnamefont\endcsname\relax
  \def\bibfnamefont#1{#1}\fi
\expandafter\ifx\csname citenamefont\endcsname\relax
  \def\citenamefont#1{#1}\fi
\expandafter\ifx\csname url\endcsname\relax
  \def\url#1{\texttt{#1}}\fi
\expandafter\ifx\csname urlprefix\endcsname\relax\def\urlprefix{URL }\fi
\providecommand{\bibinfo}[2]{#2}
\providecommand{\eprint}[2][]{\url{#2}}

\bibitem[{\citenamefont{Helstrom}(1976)}]{Helstrom1976}
\bibinfo{author}{\bibfnamefont{C.~W.} \bibnamefont{Helstrom}},
  \emph{\bibinfo{title}{{Quantum Detection and Estimation Theory}}}
  (\bibinfo{publisher}{Academic Press}, \bibinfo{address}{New York},
  \bibinfo{year}{1976}).

\bibitem[{\citenamefont{Holevo}(1973)}]{Holevo1973b}
\bibinfo{author}{\bibfnamefont{A.}~\bibnamefont{Holevo}},
  \bibinfo{journal}{Journal of Multivariate Analysis}
  \textbf{\bibinfo{volume}{3}}, \bibinfo{pages}{337} (\bibinfo{year}{1973}).

\bibitem[{\citenamefont{Yuen et~al.}(1975)\citenamefont{Yuen, Kennedy, and
  Lax}}]{Yuen1975}
\bibinfo{author}{\bibfnamefont{H.~P.} \bibnamefont{Yuen}},
  \bibinfo{author}{\bibfnamefont{R.~S.} \bibnamefont{Kennedy}},
  \bibnamefont{and} \bibinfo{author}{\bibfnamefont{M.}~\bibnamefont{Lax}},
  \bibinfo{journal}{IEEE Transactions on Information Theory}
  \textbf{\bibinfo{volume}{21}}, \bibinfo{pages}{125} (\bibinfo{year}{1975}).

\bibitem[{\citenamefont{Barnett and Croke}(2009)}]{Barnett2009}
\bibinfo{author}{\bibfnamefont{S.~M.} \bibnamefont{Barnett}} \bibnamefont{and}
  \bibinfo{author}{\bibfnamefont{S.}~\bibnamefont{Croke}},
  \bibinfo{journal}{Journal of Physics A: Mathematical and Theoretical}
  \textbf{\bibinfo{volume}{42}}, \bibinfo{pages}{062001}
  (\bibinfo{year}{2009}).

\bibitem[{\citenamefont{Konig et~al.}(2009)\citenamefont{Konig, Renner, and
  Schaffner}}]{Konig2009}
\bibinfo{author}{\bibfnamefont{R.}~\bibnamefont{Konig}},
  \bibinfo{author}{\bibfnamefont{R.}~\bibnamefont{Renner}}, \bibnamefont{and}
  \bibinfo{author}{\bibfnamefont{C.}~\bibnamefont{Schaffner}},
  \bibinfo{journal}{IEEE Transactions on Information Theory}
  \textbf{\bibinfo{volume}{55}}, \bibinfo{pages}{4337} (\bibinfo{year}{2009}).

\bibitem[{\citenamefont{Eldar et~al.}(2003)\citenamefont{Eldar, Megretski, and
  Verghese}}]{Eldar2003}
\bibinfo{author}{\bibfnamefont{Y.}~\bibnamefont{Eldar}},
  \bibinfo{author}{\bibfnamefont{A.}~\bibnamefont{Megretski}},
  \bibnamefont{and} \bibinfo{author}{\bibfnamefont{G.}~\bibnamefont{Verghese}},
  \bibinfo{journal}{IEEE Transactions on Information Theory}
  \textbf{\bibinfo{volume}{49}}, \bibinfo{pages}{1007} (\bibinfo{year}{2003}).

\bibitem[{\citenamefont{Belavkin}(1975)}]{Belavkin1975}
\bibinfo{author}{\bibfnamefont{V.~P.} \bibnamefont{Belavkin}},
  \bibinfo{journal}{Stochastics} \textbf{\bibinfo{volume}{1}},
  \bibinfo{pages}{315} (\bibinfo{year}{1975}).

\bibitem[{\citenamefont{Kennedy}(1973)}]{Kennedy1973b}
\bibinfo{author}{\bibfnamefont{R.~S.} \bibnamefont{Kennedy}},
  \bibinfo{journal}{Research Laboratory of Electronics, MIT Quarterly Progress
  Report No. 110} pp. \bibinfo{pages}{142--146} (\bibinfo{year}{1973}).

\bibitem[{\citenamefont{Hausladen and Wootters}(1994)}]{Hausladen1994}
\bibinfo{author}{\bibfnamefont{P.}~\bibnamefont{Hausladen}} \bibnamefont{and}
  \bibinfo{author}{\bibfnamefont{W.~K.} \bibnamefont{Wootters}},
  \bibinfo{journal}{Journal of Modern Optics} \textbf{\bibinfo{volume}{41}},
  \bibinfo{pages}{2385} (\bibinfo{year}{1994}).

\bibitem[{\citenamefont{Eldar and Forney}(2001)}]{Eldar2001}
\bibinfo{author}{\bibfnamefont{Y.~C.} \bibnamefont{Eldar}} \bibnamefont{and}
  \bibinfo{author}{\bibfnamefont{G.~D.~J.} \bibnamefont{Forney}},
  \bibinfo{journal}{IEEE Transactions on Information Theory}
  \textbf{\bibinfo{volume}{47}}, \bibinfo{pages}{858} (\bibinfo{year}{2001}).

\bibitem[{\citenamefont{Hausladen et~al.}(1996)\citenamefont{Hausladen, Jozsa,
  Schumacher, Westmoreland, and Wootters}}]{Hausladen1996}
\bibinfo{author}{\bibfnamefont{P.}~\bibnamefont{Hausladen}},
  \bibinfo{author}{\bibfnamefont{R.}~\bibnamefont{Jozsa}},
  \bibinfo{author}{\bibfnamefont{B.}~\bibnamefont{Schumacher}},
  \bibinfo{author}{\bibfnamefont{M.}~\bibnamefont{Westmoreland}},
  \bibnamefont{and} \bibinfo{author}{\bibfnamefont{W.~K.}
  \bibnamefont{Wootters}}, \bibinfo{journal}{Physical Review A}
  \textbf{\bibinfo{volume}{54}}, \bibinfo{pages}{1869} (\bibinfo{year}{1996}).

\bibitem[{\citenamefont{Bacon et~al.}(2005)\citenamefont{Bacon, Childs, and van
  Dam}}]{Bacon2005}
\bibinfo{author}{\bibfnamefont{D.}~\bibnamefont{Bacon}},
  \bibinfo{author}{\bibfnamefont{A.}~\bibnamefont{Childs}}, \bibnamefont{and}
  \bibinfo{author}{\bibfnamefont{W.}~\bibnamefont{van Dam}}, in
  \emph{\bibinfo{booktitle}{46th Annual IEEE Symposium on Foundations of
  Computer Science (FOCS'05)}} (\bibinfo{publisher}{IEEE},
  \bibinfo{year}{2005}), pp. \bibinfo{pages}{469--478}.

\bibitem[{\citenamefont{Moore and Russell}(2007)}]{Moore2007}
\bibinfo{author}{\bibfnamefont{C.}~\bibnamefont{Moore}} \bibnamefont{and}
  \bibinfo{author}{\bibfnamefont{A.}~\bibnamefont{Russell}},
  \bibinfo{journal}{Quantum Information \& Computation}
  \textbf{\bibinfo{volume}{7}}, \bibinfo{pages}{752} (\bibinfo{year}{2007}).

\bibitem[{\citenamefont{Hayashi et~al.}(2008)\citenamefont{Hayashi, Kawachi,
  and Kobayashi}}]{Hayashi2008}
\bibinfo{author}{\bibfnamefont{M.}~\bibnamefont{Hayashi}},
  \bibinfo{author}{\bibfnamefont{A.}~\bibnamefont{Kawachi}}, \bibnamefont{and}
  \bibinfo{author}{\bibfnamefont{H.}~\bibnamefont{Kobayashi}},
  \bibinfo{journal}{Quantum Information \& Computation}
  \textbf{\bibinfo{volume}{8}}, \bibinfo{pages}{345} (\bibinfo{year}{2008}).

\bibitem[{\citenamefont{Ban et~al.}(1997)\citenamefont{Ban, Kurokawa, Momose,
  and Hirota}}]{Ban1997}
\bibinfo{author}{\bibfnamefont{M.}~\bibnamefont{Ban}},
  \bibinfo{author}{\bibfnamefont{K.}~\bibnamefont{Kurokawa}},
  \bibinfo{author}{\bibfnamefont{R.}~\bibnamefont{Momose}}, \bibnamefont{and}
  \bibinfo{author}{\bibfnamefont{O.}~\bibnamefont{Hirota}},
  \bibinfo{journal}{International Journal of Theoretical Physics}
  \textbf{\bibinfo{volume}{36}}, \bibinfo{pages}{1269} (\bibinfo{year}{1997}).

\bibitem[{\citenamefont{Kato et~al.}(1999)\citenamefont{Kato, Osaki, Sasaki,
  and Hirota}}]{Kato1999}
\bibinfo{author}{\bibfnamefont{K.}~\bibnamefont{Kato}},
  \bibinfo{author}{\bibfnamefont{M.}~\bibnamefont{Osaki}},
  \bibinfo{author}{\bibfnamefont{M.}~\bibnamefont{Sasaki}}, \bibnamefont{and}
  \bibinfo{author}{\bibfnamefont{O.}~\bibnamefont{Hirota}},
  \bibinfo{journal}{IEEE Transactions on Communications}
  \textbf{\bibinfo{volume}{47}}, \bibinfo{pages}{248} (\bibinfo{year}{1999}).

\bibitem[{\citenamefont{Cariolaro and
  Pierobon}(2010{\natexlab{a}})}]{Cariolaro2010}
\bibinfo{author}{\bibfnamefont{G.}~\bibnamefont{Cariolaro}} \bibnamefont{and}
  \bibinfo{author}{\bibfnamefont{G.}~\bibnamefont{Pierobon}},
  \bibinfo{journal}{IEEE Transactions on Communications}
  \textbf{\bibinfo{volume}{58}}, \bibinfo{pages}{1213}
  (\bibinfo{year}{2010}{\natexlab{a}}).

\bibitem[{\citenamefont{Eldar et~al.}(2004)\citenamefont{Eldar, Megretski, and
  Verghese}}]{Eldar2004}
\bibinfo{author}{\bibfnamefont{Y.~C.} \bibnamefont{Eldar}},
  \bibinfo{author}{\bibfnamefont{A.}~\bibnamefont{Megretski}},
  \bibnamefont{and} \bibinfo{author}{\bibfnamefont{G.}~\bibnamefont{Verghese}},
  \bibinfo{journal}{IEEE Transactions on Information Theory}
  \textbf{\bibinfo{volume}{50}}, \bibinfo{pages}{1198} (\bibinfo{year}{2004}).

\bibitem[{\citenamefont{Cariolaro and
  Pierobon}(2010{\natexlab{b}})}]{Cariolaro2010b}
\bibinfo{author}{\bibfnamefont{G.}~\bibnamefont{Cariolaro}} \bibnamefont{and}
  \bibinfo{author}{\bibfnamefont{G.}~\bibnamefont{Pierobon}},
  \bibinfo{journal}{IEEE Transactions on Communications}
  \textbf{\bibinfo{volume}{58}}, \bibinfo{pages}{623}
  (\bibinfo{year}{2010}{\natexlab{b}}).

\bibitem[{\citenamefont{Helstrom}(1982)}]{Helstrom1982}
\bibinfo{author}{\bibfnamefont{C.}~\bibnamefont{Helstrom}},
  \bibinfo{journal}{IEEE Transactions on Information Theory}
  \textbf{\bibinfo{volume}{28}}, \bibinfo{pages}{359} (\bibinfo{year}{1982}).

\bibitem[{\citenamefont{Sasaki et~al.}(1998)\citenamefont{Sasaki, Kato, Izutsu,
  and Hirota}}]{Sasaki1998}
\bibinfo{author}{\bibfnamefont{M.}~\bibnamefont{Sasaki}},
  \bibinfo{author}{\bibfnamefont{K.}~\bibnamefont{Kato}},
  \bibinfo{author}{\bibfnamefont{M.}~\bibnamefont{Izutsu}}, \bibnamefont{and}
  \bibinfo{author}{\bibfnamefont{O.}~\bibnamefont{Hirota}},
  \bibinfo{journal}{Physical Review A} \textbf{\bibinfo{volume}{58}},
  \bibinfo{pages}{146} (\bibinfo{year}{1998}).

\bibitem[{\citenamefont{Horn and Johnson}(1990)}]{HornJohnson}
\bibinfo{author}{\bibfnamefont{R.~A.} \bibnamefont{Horn}} \bibnamefont{and}
  \bibinfo{author}{\bibfnamefont{C.~R.} \bibnamefont{Johnson}},
  \emph{\bibinfo{title}{{Matrix Analysis}}} (\bibinfo{publisher}{Cambridge
  University Press}, \bibinfo{address}{New York}, \bibinfo{year}{1990}).

\bibitem[{\citenamefont{Davis}(1979)}]{Davis1979}
\bibinfo{author}{\bibfnamefont{P.~J.} \bibnamefont{Davis}},
  \emph{\bibinfo{title}{{Circulant Matrices: Second Edition}}}, AMS Chelsea
  Publishing (\bibinfo{publisher}{Wiley}, \bibinfo{address}{New York},
  \bibinfo{year}{1979}).

\bibitem[{\citenamefont{Mochon}(2006)}]{Mochon2006}
\bibinfo{author}{\bibfnamefont{C.}~\bibnamefont{Mochon}},
  \bibinfo{journal}{Phys. Rev. A} \textbf{\bibinfo{volume}{73}},
  \bibinfo{pages}{032328} (\bibinfo{year}{2006}).

\bibitem[{\citenamefont{Kato}(2013)}]{Kato2013}
\bibinfo{author}{\bibfnamefont{K.}~\bibnamefont{Kato}},
  \bibinfo{journal}{Tamagawa University Quantum ICT Research Institute
  Bulletin} \textbf{\bibinfo{volume}{3}}, \bibinfo{pages}{29}
  (\bibinfo{year}{2013}).

\bibitem[{\citenamefont{Dolinar et~al.}(2006)\citenamefont{Dolinar, Hamkins,
  Moision, and Vilnrotter}}]{DescansoCap4}
\bibinfo{author}{\bibfnamefont{S.}~\bibnamefont{Dolinar}},
  \bibinfo{author}{\bibfnamefont{J.}~\bibnamefont{Hamkins}},
  \bibinfo{author}{\bibfnamefont{B.}~\bibnamefont{Moision}}, \bibnamefont{and}
  \bibinfo{author}{\bibfnamefont{V.}~\bibnamefont{Vilnrotter}}, in
  \emph{\bibinfo{booktitle}{Deep Space Optical Communications}}, edited by
  \bibinfo{editor}{\bibfnamefont{H.}~\bibnamefont{Hemmati}}
  (\bibinfo{publisher}{Wiley}, \bibinfo{address}{Hoboken, New Jersey},
  \bibinfo{year}{2006}), chap.~\bibinfo{chapter}{4}.

\end{thebibliography}

\end{document}